\documentclass{aa}
\usepackage{graphicx}
\usepackage{txfonts}
\usepackage{natbib}
\usepackage{here}
\usepackage[normalem]{ulem}

\newcommand\be{\begin{equation}}
\newcommand\ee{\end{equation}}
\newcommand{\ammonia}{\mbox{NH$_3$}}

\newcommand\micron{\mbox{$\mu$m}}

\newcommand\arcdeg{\mbox{$^\circ$}}%
\newcommand\bea{\begin{eqnarray}}
\newcommand\eea{\end{eqnarray}}
\newcommand\HII{\hbox{H\,{\sc ii}}}

\newcommand{\msun}{M$_{\rm \odot}$}

\newcommand{\thCO}{$^{13}$CO}

\newcommand{\kms}{\mbox{kms$^{-1}$}}

\newcommand{\tdust}{\mbox{$T_{\rm d}$}}
\newcommand{\lsun}{\ensuremath{\mathrm{L}_\odot}}
\def\phn{\phantom{0}}
\newcommand{\ea}{et al.}

\begin{document}

\title{SCUBA Mapping of Outer Galaxy Protostellar Candidates}

\author{B. Mookerjea\inst{1,2}, G.  Sandell\inst{3}, J.
Stutzki\inst{1}, J. G. A.  Wouterloot\inst{4}, }

\offprints{bhaswati@tifr.res.in}
\institute{KOSMA, I. Physikalisches Institut, Universit\"at zu K\"oln,
Z\"ulpicher Strasse 77, 50937 Cologne, Germany
\and
Department of Astronomy, University of Maryland, College
Park, MD 20742, USA
\and
SOFIA-USRA, NASA Ames Research Center, MS N211-3, Moffett Field, CA 94035
\and 
Joint Astronomy Centre, 660 N. A'ohoku Place, University Park,
Hilo, Hawaii 96720 , USA }

\date{Accepted . Received ; in original form }
\authorrunning{Mookerjea et al.}
\titlerunning{SCUBA studies of Outer Galaxy IRAS sources}


\abstract
{}
{We aim to study dust properties of massive star forming regions in
the outer Galaxy, in a direction opposite to the Galactic center.}
{We present observations of six outer Galaxy point sources IRAS
01045+6505, 01420+6401, 05271+3059, 05345+3556, 20222+3541 and
20406+4555, taken with the Submillimeter Common-User Bolometer Array
(SCUBA) on the James Clerk Maxwell Telescope (JCMT) at 450 and
850~\micron. Single temperature greybody models are fitted to the
Spectral Energy Distribution of the detected sub-mm cores to derive dust
temperature, dust emissivity index and optical depth at 250~\micron. The
observed radial intensity profiles of the sub-mm cores were fitted with
power laws to derive the indices describing the density
distribution. }
{At a resolution of 15\arcsec\ all six IRAS point sources show multiple
emission peaks. Only four out of fourteen detected sub-mm cores show
associated mid-infrared emission. For the sub-mm cores we derive dust
temperatures of 32$\pm$5~K and dust emissivity indices between 0.9 and
2.5. The density profiles of the sub-mm cores can be fitted by a
single power law distribution with indices -1.5$\pm$0.3, with most
cores showing an index of -1.5. This is consistent with most
observations of massive star forming regions and supports predictions
of models of star formation which consider non-thermal support against
gravitational collapse.}{}

\keywords
{stars: formation -- ISM:general -- ISM:\HII\ regions --
ISM:dust,extinction -- submillimeter -- ISM:individual: IRAS 05271+3059,
05345+3556, 20222+3541, 20406+4555.}
\maketitle

\section{Introduction}

With the development of high angular resolution millimeter (mm) and
sub-millimeter (sub-mm) instrumentation, observations studying the
influence of different physical environment on the formation of
molecular clouds subsequently developing into protostellar cores has
been receiving increased attention. While the inner Galaxy has been
extensively studied, lack of instruments with high angular resolution
has limited the study of sources in the outer Galactic disk which are
in a potentially different physical and chemical environment. Existing
observations suggest that in the outer Galactic disk the molecular
clouds are more sparsely distributed \citep{wouterloot1990}, the
diffuse galactic interstellar radiation field is weaker
\citep{cox1989,bloemen1985}, the metallicity is lower
\citep{shaver1983,fich1991,rudolph2006} and the cosmic-ray flux
density is smaller \citep{bloemen1984}, compared to the solar
neighbourhood. Although the efficiency at which molecular clouds form
from atomic gas is much lower in the far outer Galaxy, the mass
spectrum of molecular clumps is not significantly different from the
inner Galaxy \citep{snell2002}.  The spectra however show a tendency
towards steepening in the far outer Galaxy.  The star formation
activity in the outer Galaxy clouds is also similar to the inner
Galaxy clouds. Thus the observed low global star formation rate in the
outer Galaxy is a consequence of the inefficiency of the formation of
molecular clouds in this region and not an inefficiency in the star
formation process within these clouds \citep{snell2002}. The outer
Galaxy star forming regions could thus be better examples of stars
and/or stellar clusters forming in isolation in contrast to the
crowded fields within the inner Galaxy.  It is thus important to
study the point sources identified by IRAS in the outer Galactic disk
at high angular resolutions to probe their multiplicity, dust
properties and kinematics.  As a first step towards these goals we
present here sub-mm continuum mapping of selected outer Galaxy IRAS
sources.


We have selected a sample of outer Galaxy (with Galactocentric
distances greater than 10~kpc) sources from the IRAS point source
catalog (PSC) based on the following criteria: F$_{100\micron}\geq
100$~Jy, F$_{100\micron}>$F$_{60\micron}$, 30\arcdeg$\leq
l\leq$330\arcdeg, -5\arcdeg$\leq b\leq$5\arcdeg. The Galactic
longitude criterion ensures that source confusion along the line of
sight faced towards the inner Galaxy, is avoided.  The sample consists
of 17 sources, all of which are {\em bona-fide} point sources
according to the IRAS PSC. For majority of these sources no continuum
maps other than the IRAS images are available.  Here we present sub-mm
(450 and 850~\micron) observations of six sources in our sample, {\em
viz.,} IRAS 01045+6505, IRAS 01420+6401, IRAS 05271+3059, 05345+3556,
20222+3541, and 20406+4555. For the sources IRAS 01045+6505 and IRAS
01420+6401 we have only 850~\micron\ observations, hence in this
paper, we present only the basic results for these two sources  and
discuss the other four sources in detail.

The purpose of these observations was primarily to obtain a higher
angular resolution view in order to detect the sub-structures within
these sources. Using these observations we also derive the dust
properties of these regions in order to improve our understanding of
the true nature of sources located in the outer Galaxy environments. 


\section{Observations and data reduction}
\label{sec_obs}

We present here results of observations made between September 2004
and January 2005, using the Submillimeter Common-User Bolometer Array
(SCUBA) on the James Clerk Maxwell Telescope \citep{holland1999}. The
SCUBA array covers a hexagonal 2\farcm5 field of view with 91 and 37
pixels at 450 and 850~\micron\ respectively. Maps are fully sampled
using the ``jiggle" mode, in which the telescope is moved around a
64-position pattern by the secondary mirror in order to fully sample
the sky with the wider intrinsic beam spacing of the array. A chop
throw of 120\arcsec\ in azimuth was used for all observations. 


The data were reduced using the SCUBA User Reduction Facility
\citep[SURF;][]{jenness1999} and the STARLINK imaging software
following the methods described in the SCUBA mapping cookbook
\citep{sandell2001}. The steps of reduction included flat-fielding,
extinction correction, sky-subtraction and calibration of images in
Jy~beam$^{-1}$.  The maps were calibrated using observations of
primary calibrators like Uranus, CRL~618 and CRL~2688 observed close
to the time of our observations.
 
All the maps were converted to FITS-files and exported to MIRIAD
\citep{sault1995} for further analysis. In order to correct for the
error lobe contribution, we deconvolved all the maps using CLEAN and a
symmetry-sized model beam.  The model beam, derived from the 450 and
850~\micron\ observations of Uranus obtained during the same night and
under similar conditions, and consistent with repeated observations
during the whole observing run, is composed of three symmetric
Gaussians.  At 850~\micron\ it has HPBWs of 15\farcs2, 58\farcs6 and
140\arcsec, with amplitudes 0.985, 0.013 and 0.002, respectively. At
450~\micron\ it has HPBWs of 8\farcs5, 33\arcsec\ and 140\arcsec\ with
amplitudes 0.942, 0.055 and 0.003 respectively.  In order to obtain
images with good S/N we have restored maps at both 450 and
850~\micron\ to a common resolution of 15\arcsec.

\section{Results} 
\label{sec_results}

\begin{figure*}
\includegraphics[width=16.0cm,angle=0]{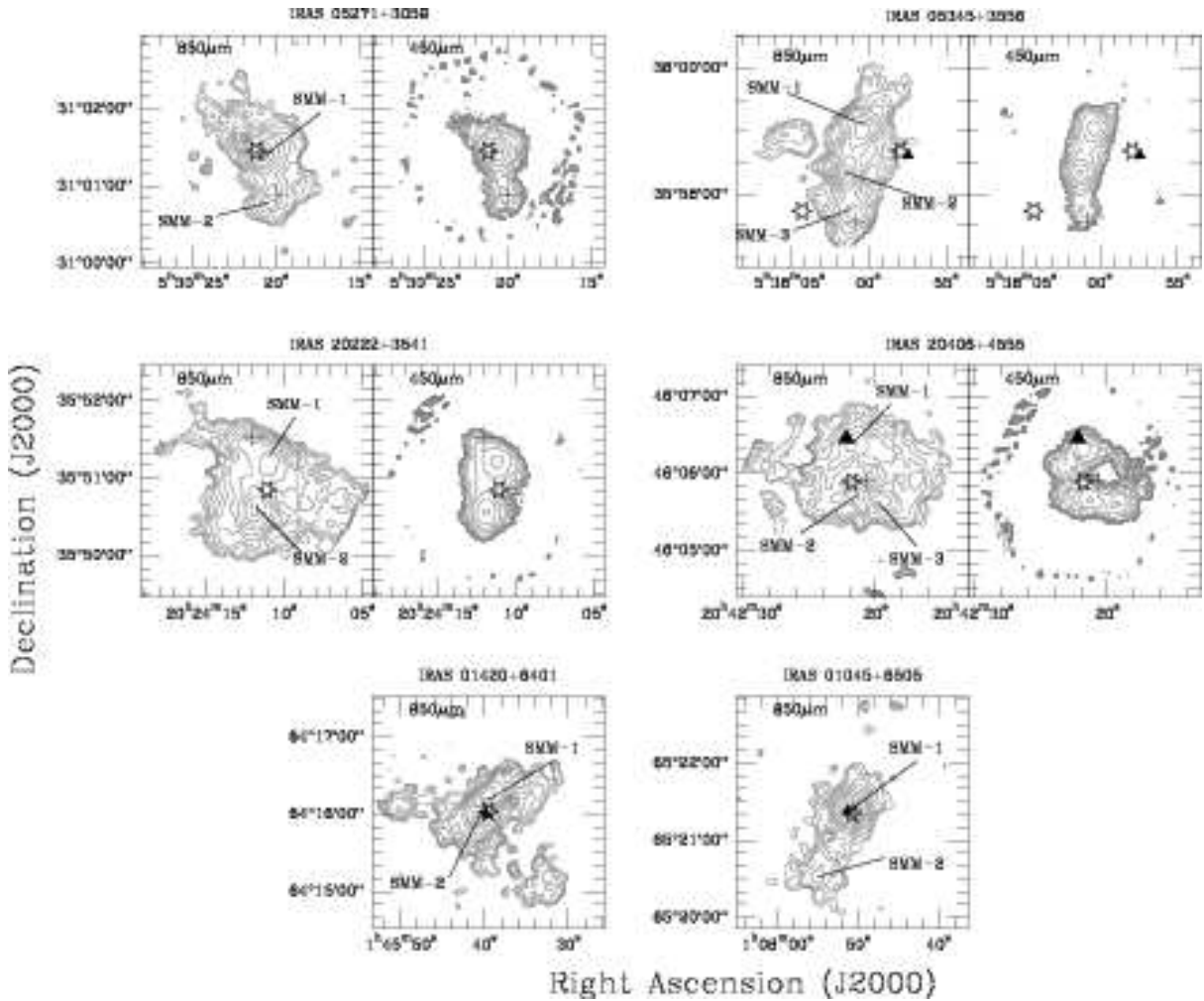}
\caption{Contours of  450 \& 850~\micron\ emission for the  outer
Galaxy sources. `+' and $\ast$ indicate positions of MSX and IRAS
point sources respectively in the mapped regions. Filled triangles
show locations of radio continuum sources.  The peak flux densities
(rms) are as follows: (1) 05271+3059: F$_{850}$ = 0.6 (0.02) Jy/beam,
F$_{450}$ = 4.7 (0.03) Jy/beam; (2) 05345+3556: F$_{850}$ = 1.0 (0.03)
Jy/beam, F$_{450}$ = 8.0 (0.04) Jy/beam; (3) 20222+3541: F$_{850}$ =
0.5 (0.02) Jy/beam, F$_{450}$ = 3.2 (0.02) Jy/beam; (4) 20406+4555:
F$_{850}$ = 0.5 (0.02) Jy/beam F$_{450}$ = 1.8 (0.01) Jy/beam; (5)
01420+6401: F$_{850}$ = 0.4 (0.02); (6) 01045+6505: F$_{850}$ = 1.2
(0.03) Jy/beam.  11 equally spaced logarithmic contours are drawn,
with the lowest level at 2$\sigma$ and the highest level at 95\% of
the peak value.
\label{fig_intmap}}
\end{figure*}

Figure~\ref{fig_intmap} shows the 450 and 850~\micron\ dust continuum
maps of the six outer Galaxy IRAS point sources.  We find that all the
six sources are resolved into at least two to three emission peaks in
the sub-mm images.  Comparison with the MSX point source catalog
suggests that most of the sub-mm cores do not have any associated MSX
source. Given the 4\arcsec\ positional accuracy of the SCUBA maps (see
below), we consider a sub-mm core to be associated with an MSX source
only if the MSX source is within 5\arcsec\ of the position of the
sub-mm core.   Given the large distances to the sources,
displacement of the young star from its core due to proper motion
outside of this association range is rather unlikely.  Our
interpretation is that the lack of MSX counterparts for the sub-mm
sources is due to differences in the evolutionary stages. Based on
this criterion we find that only 05271-SMM1, 05271-SMM2 and 01045-SMM1
have associated MSX sources.  Only IRAS 01045+6505 was observed using
IRAC and MIPS instruments on Spitzer.  Preliminary results from
Spitzer show that both 01045-SMM1 and 01045-SMM2 have sources at
24~\micron\ associated with them, with the IR source at the position
of 01045-SMM1 being partially saturated.  This indicates that
01045-SMM1 is a reasonably hot source with a dust temperature of 40~K
or more.  IRAS 20406+4555, 01420+6401 and 01045+6505 have associated
radio continuum sources.  For IRAS 01420+6401 and 01045+6505 the radio
continuum positions are based on high resolution VLA observations
\citep{rudolph1996}, whereas for IRAS 20406+4555 the radio data is
from the GB6 catalog \citep{gregory1996}.  The nominal IRAS position
for 05345+3556 is offset with respect to the ridge detected in the
sub-mm.  We have checked the ISSA images for this source and find that
at the position given by the IRAS PSC there is no prominent peak.  In
Sect.~\ref{sec_targets} we discuss the sources individually in the
light of the available limited complementary observations.

Given the extended nature of these sources and the 2\arcmin\ chop throw,
the observed off positions will not have been completely free of
emission, and this will have been subtracted from the fluxes and as a
background from the images. However, as we are more interested in the
warmer dust corresponding to the warm cores seen by IRAS, rather than
the cold diffuse component, the background-subtracted flux measurements
are appropriate.  In order to derive the parameters of the sub-mm cores
detected in the IRAS sources, we have fitted an elliptical Gaussian  for
the sub-mm source, and a background plane for the surrounding cloud
using the task IMFIT in MIRIAD.  The fit to the broader component is
mainly to provide a good subtraction of the extended emission, and is
not to estimate the flux density of the surrounding cloud.
Table~\ref{tab_phot} presents results of the fits to the detected
sub-mm sources. We estimate the uncertainty in determining the positions
of the sub-mm sources based on the Gaussian fitting  to be 1\arcsec.
Combining this with the pointing accuracy of the observations
($<3$\arcsec) we estimate the final astrometric uncertainty of the
positions of the sub-mm cores to be $\sim 4$\arcsec. The sizes are
derived from the Gaussian fits to the 450~\micron\ data and the flux
densities at 850~\micron\ are derived by constraining the sizes of the
Gaussian fits to be the same as the sizes at 450~\micron.  For the sub-mm
cores not clearly detected at 450~\micron, we have derived the sizes
from the 850~\micron\ map and integrated the 450~\micron\ fluxes over an
aperture of same size.

\begin{table*}
\begin{center}
\caption{Observed properties of the sub-mm sources identified 
in the SCUBA maps
\label{tab_phot}}
\begin{tabular}{cllcccccc}
\hline \hline
Submillimeter & $\alpha (2000)$ & $\delta (2000)$ &
D &$\theta_a\times\theta_b$$^a$&
S$_{450}$ & S$_{850}$ & IR & Radio\\
Source& & &
kpc & (\arcsec$\times$\arcsec)&Jy&Jy & Assoc. & Assoc.\\
\hline
&&&&&&&\\
05271+3059 SMM-1 & 05:30:20.97 &  31:01:25.8 & 16.5 & $12.9\times12.8$ & 
7.92 & 1.08 & MSX & \ldots\\
05271+3059 SMM-2 & 05:30:20.28 &  31:00:50.3 & 16.5 & $15.1\times12.7$ & 2.26
& 0.24  & MSX & \ldots\\
&&&&&&&\\
\hline
&&&&&&&\\
05345+3556 SMM-1 & 05:38:00.41 &  35:59:00.2 & 14.0 & $19.2\times14.1$ & 14.18
& 1.65 & \ldots & \ldots \\
05345+3556 SMM-2 & 05:38:01.29  &  35:58:21.2 & 14.0 & $13.9\times\phn9.2$ & 
12.74 & 1.14 & \ldots & \ldots \\
05345+3556 SMM-3 & 05:38:00.99  &  35:57:45.7 &14.0 &  $17.7\times14.5$
&$^b$& 0.80 & \ldots & \ldots \\
&&&&&&&\\
\hline
&&&&&&&\\
20222+3541 SMM-1 & 20:24:10.69 &  35:51:10.2 & 2.2 & $20.4\times20.0$ & \phn6.06
& 0.33 & \ldots & \ldots \\
20222+3541 SMM-2 & 20:24:11.85 &  35:50:34.8 & 2.2 & $15.0\times13.2$ & \phn5.95
& 0.51 & \ldots & \ldots \\
&&&&&&&\\
\hline
&&&&&&&\\
20406+4555 SMM-1 & 20:42:21.75 & 46:06:21.1 & 7.8 & $15.4\times11.0$ & \phn1.79  
& 0.17 & \ldots & GB6 \\
20406+4555 SMM-2 & 20:42:21.30 & 46:05:42.8 & 7.8 & $12.7\times12.6$ & \phn3.02 
& 0.30 & \ldots & \ldots \\
20406+4555 SMM-3 & 20:42:19.33 & 46:05:38.3 & 7.8 & $15.7\times11.3$ & \phn2.17
& 0.24 & \ldots & \ldots \\
&&&&&&&\\
\hline
&&&&&&&\\
01420+6401-SMM1 & 01:45:39.09 & 64:16:15.9 & 10.1 & 14.8$\times$10.8 &
\ldots & 0.44 & \ldots & GB6,VLA\\
01420+6401-SMM2 & 01:45:40.76 & 64:15:57.3 & 10.1 & 16.5$\times$16.3 &
\ldots & 0.71 & \ldots & \ldots\\
&&&&&&&\\
\hline
&&&&&&&\\
01045+6505-SMM1 & 01:07:50.99 & 65:21:23.2 & 11.3 & $12.7\times11.6$ &
\ldots & 1.81  & MSX,Spitzer & GB6,VLA\\
01045+6505-SMM2 & 01:07:54.99 & 65:20:31.7 & 11.3 & $16.7\times16.5$ &
\ldots & 0.46  & Spitzer & \ldots\\
&&&&&&&\\
\hline
\end{tabular}
\flushleft
$^{a}$ Deconvolved Major ($\theta_a$) and Minor ($\theta_b$) axes
derived from elliptical Gaussian fits.\\
$^b$
Not detected at 450~\micron. $S_{450}$ obtained by integrating at the
position of the 850~\micron\ peak is 5.29~Jy.\\
\end{center}
\end{table*}

\section{Notes on the sources}
\label{sec_targets}

Here we present details of existing observations of the four sources
IRAS 05271+3059, 05345+3556, 20222+3541 and 20406+4555.

{\bf IRAS 05271+3059:} This source, WB89~651, is  part of the outer
Galaxy catalog compiled by \citet{wouterloot1989}. Based on the
velocity information from the molecular line observations the distance
to the source is estimated to be 16.5~kpc.  The luminosity of the
source estimated from the 60 and 100~\micron\ IRAS fluxes is
4.2~10$^4$~\lsun.  CS(2--1) \citep{bronfman1996} and H$_2$O
\citep{valdettaro2001} maser emission was found to be associated with
this source, although no methanol maser was detected
\citep{slysh1999}.  There is no radio continuum source known to be
associated with IRAS 05271+3058.  We find that both sub-mm sources are
associated with MSX point sources (Fig.~\ref{fig_intmap}).

{\bf IRAS 05345+3556:} Also known as WB89~673, this source is at a
distance of 14~kpc as derived from molecular line observations by
\citet{wouterloot1989} and has a  luminosity of 1.7~10$^4$~\lsun. The
radio continuum source detected in NRAO VLA Sky Survey (NVSS) at
1.4~GHz \citep{condon1998} is coincident with the nominal position of
the infrared peak according to the IRAS Point Source Catalog. As noted
in Sect.~\ref{sec_results} the dust emission detected in the SCUBA maps
is along a ridge which is offset from this nominal infrared peak
position. Similar to IRAS 05271+3059, CS(2--1) \citep{bronfman1996},
SiO \citep{harju1998} and H$_2$O \citep{valdettaro2001} maser
emissions are found to be associated with this source and no methanol
was detected \citep{slysh1999}. No MSX point sources were found within
a search radius of 20\arcsec\ around the sources SMM-1 and SMM-2, only
SMM-3 has a nearby MSX source.

{\bf IRAS 20222+3541:} \citet{beck1988} suggested that IRAS
20222+3541, also known as MCG 06-45-001, is a local group galaxy. In a
subsequent paper, \citet{shore1990} used CO observations to
conclusively show that rather than being an external galaxy, this IRAS
source is a Galactic star forming region with an associated \HII\
region. Based on the CO observations of \citet{shore1990} we estimate
the kinematic distance to the source to be 2.2~kpc  for a v$_{\rm
lsr}$ of $\sim$7.4 km/s. This is the nearest of the sources in our
sample. The luminosity of the source is 1.5~10$^{3}$~\lsun.  Only one
point source from the MSX catalog is found in the mapped region 
although  the images of almost all MSX bands show an
extended emission coincident with the IRAS source. The only identified
MSX point source in this region is offset from SMM-1 by $\sim
25$\arcsec.  In contrast to \citet{shore1990} we find that the only
radio continuum source detected in the NRAO VLA Sky Survey (NVSS) at
1.4~GHz \citep{condon1998} is offset by $\sim 2$\arcmin\ with respect
to the dust continuum peaks seen in IRAS, MSX and SCUBA datasets. In

{\bf IRAS 20406+4555:} This source, also known as WB89~4 from the
catalog of outer Galaxy sources by \citet{wouterloot1989}, is located
at a distance of 7.8~kpc. We have confirmed from our recently concluded
and yet unpublished CO observations, as well as from the \thCO\ J=1--0
observations of \citet{wu2001} that only the velocity component at
$-49$~\kms\ is associated with this source and this corresponds to the
aforementioned distance. The luminosity of the source based on the IRAS
fluxes is 2.6~10$^{4}$~\lsun. There are no MSX point sources in the
MSX6C associated with this region, although both in bands A and E the
source is clearly detected.  An \HII\ region from the GB6 catalog of
radio sources \citep{gregory1996}  is centerd at the position of SMM-1,
while the IRAS source is located closer to SMM-2. The comparatively
compact appearance of SMM-2 \& 3 together with the \HII\ region might
suggest that SMM-2/3 contain protostars, formation of which is triggered
by the \HII\ region. However so far no \ammonia\ and H$_2$O maser
emission were detected from this source \citep{molinari1996}.

\section{Spectral Energy Distributions of the sub-mm cores}

Several authors \citep{sridharan2002,beuther2002} have shown that two
temperature components are present in the SED of luminous YSOs: a
compact hot component which dominates the IRAS 12~\micron\  and
25~\micron\ fluxes, and a more extended component arising from colder
gas which dominates the IRAS 60~\micron\ and 100~\micron\ fluxes.
Here we analyze the spectral energy distributions (SED) of the sub-mm
cores detected at both SCUBA wavelengths, between 60 and 850~\micron\
in order to derive order of magnitude estimates of the dust
temperature and mass of the cold component.  Since the sub-mm cores
are not resolved in the mid and far-infrared, we apportion the 60 and
100~\micron\ flux densities for the IRAS sources to the sub-mm cores
in the same ratio as the observed 450~\micron\ fluxes.

The greybody functions have the form \citep[e.g.,][]{dent1998},

\begin{equation}
F_\nu = \Omega B_\nu(T_{dust})(1-e^{-\tau_\nu}),
\end{equation}

where F$_\nu$ is the flux measured at a frequency $\nu$, $\Omega$ is
the solid angle subtended by the cloud, B$_\nu(T_{dust})$ is the
Planck function evaluated at the dust temperature \tdust and $\nu$,
and $\tau_\nu$ is the optical depth at $\nu$. The optical depth
$\tau_\nu$ scales as $\tau_0 \nu^\beta$, where $\beta$ is the grain
emissivity index and $\tau_0$ is the optical depth at a reference
frequency ($\nu_0$).

Fitting was performed such that $\chi^2$ was minimized, with $\chi^2$
defined as,

\begin{equation}
\chi^2  = \sum
\left[1-\left(\frac{F_{\nu,model}}{F_{\nu,obs}}\right)\right]^2
\end{equation}

This definition of $\chi^2$ gives equal weighting to the different
wavelength regimes \citep{hunter2000}. We have constrained the angular
sizes of the sub-mm cores to those measured in the SCUBA observations
(Table~\ref{greytab}). Three free parameters were used: the
temperature, the optical depth at 250~\micron\ and the grain
emissivity index ($\beta$).

\begin{figure}
\includegraphics[width=8cm,angle=0]{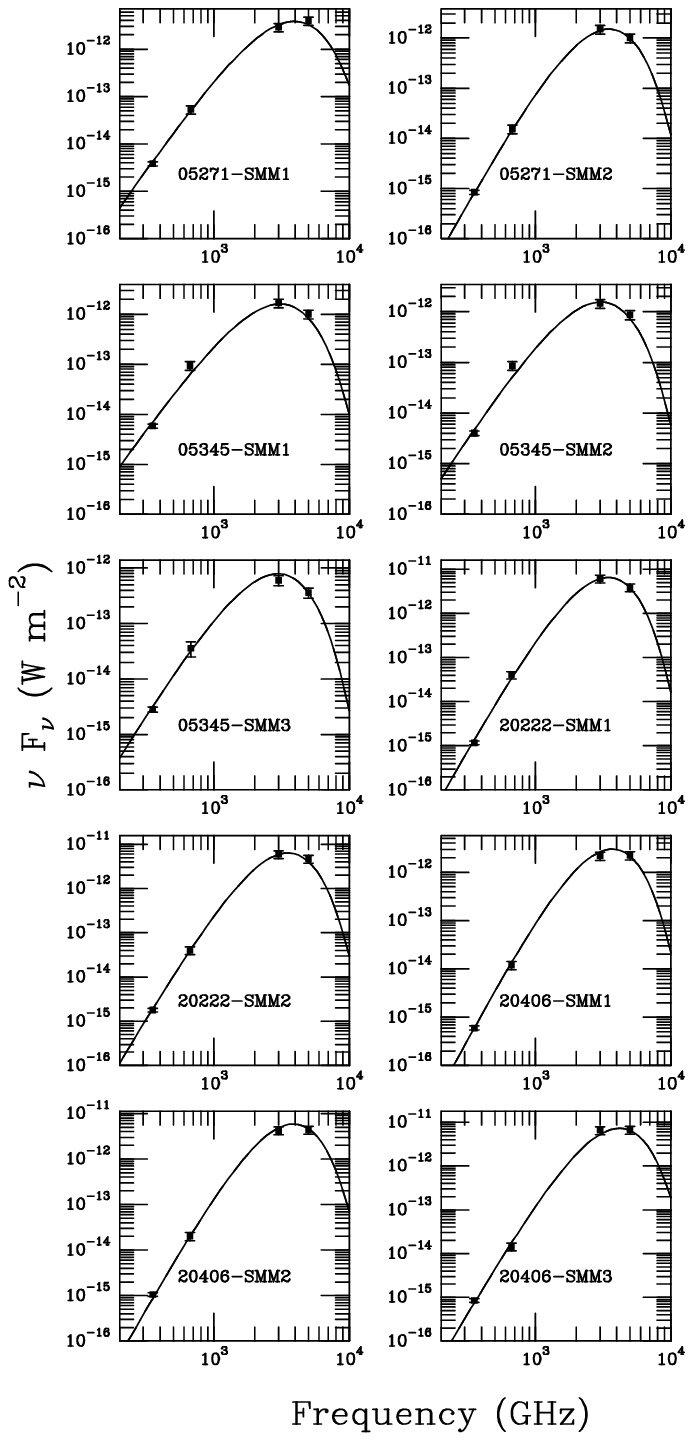}
\caption{Single temperature greybody fits to the Spectral Energy
Distribution (SED) of sub-mm cores. Fitted parameters are presented in
Table~\ref{greytab}\label{greyfit}}
\end{figure}

Figure~\ref{greyfit} shows the greybody models fitted to the SEDs and
Table~\ref{greytab} also presents the parameters like mass, luminosity
etc. calculated for the best fit model. The mass has been estimated
from the fitted values of $\tau_{250}$, \tdust\ and $\beta$,  adopting
the ``Hildebrand'' mass opacity, ${\rm \kappa_o}$, defined at 250
$\mu$m (1200 GHz), i.e.  $\kappa_{\rm 1200GHz}$ = 0.1 cm$^{2}$g$^{-1}$
\citep{hildebrand1983} and a gas-to-dust ratio of 100. The main
uncertainty in calculating the dust masses here lie in the use of
inner Galaxy dust properties as well as gas-to-dust ratio for these
outer Galaxy sources.  Observationally there is evidence of reduced
metallicity towards the outer Galaxy \citep{rudolph2006} and the
dust-to-gas ratio depends almost linearly on metallicity
\citep{boselli2002}. The assumed gas-to-dust ratio of 100 does not
take the metallicity gradient into account. In addition, the
calculated dust mass opacities depend on the grain composition and
even for inner Galaxy sources the commonly used values differ by as
much as factors of 3.  Overall, we estimate that the calculated mass
can deviate from the real value by up to a factor of 10. For the
sub-mm sources in IRAS 01045+6505 and 01420+6401 in the absence of the
450~\micron\ fluxes greybody SED fitting was not possible. We have
estimated the mass of these sub-mm cores assuming \tdust\ to be 32~K,
average value of \tdust\ found from the SED fits and $\beta=2$.  As
discussed in Sect.~\ref{sec_dustprop} a choice of $\beta=2$ can be
reasonably justified based on both theoretical and observational
results. 

%


\begin{table*}
\caption{Parameters of the SED Models for the sub-mm cores
\label{greytab}}
\begin{tabular}{llccccc}
\hline
Source & Dia & T & $\tau_{250}$& $\beta$ & L & M \\
 & pc & K & & & \lsun & \msun\\
\hline
\hline
05271-SMM1 &  1.03 & 36.9 & 0.01 & 1.2 & 42272 & \phn711\\
05271-SMM2 &  1.10 & 28.6 & 0.01 & 1.9 & 15184 & \phn814\\
&&&&&&\\
05345-SMM1 &  1.12 & 30.7 & 0.01 & 0.9 & 13192 & \phn595\\
05345-SMM2 &  0.77 & 28.6 & 0.03 & 1.2 & 12270 & \phn778\\
05345-SMM3 &  1.09 & 28.6 & 0.01 & 1.0 & \phn6477 & \phn398\\
&&&&&&\\
20222-SMM1 &  0.22 & 26.6 & 0.02 & 2.5 & 1091  & \phn\phn86 \\
20222-SMM2 &  0.15 & 28.6 & 0.03 & 2.2 & 1083  & \phn\phn61 \\
&&&&&&\\
20406-SMM1 &  0.49 & 28.6 & 0.01 & 2.4 & 6448  & \phn320 \\
20406-SMM2 &  0.48 & 30.7 & 0.02 & 2.4 & 12463 & \phn422 \\
20406-SMM3 &  0.50 & 32.8 & 0.01 & 2.4 & 15387 & \phn334 \\
&&&&&&\\
01420-SMM1 &  \ldots & 32$^{a}$ & \ldots  & \dots & \ldots & 1041$^{b}$ \\
01420-SMM2 &  \ldots & 32$^{a}$ & \ldots  & \dots & \ldots & \phn350$^{b}$ \\
&&&&&&\\
01045-SMM1 &  \ldots & 40$^{a}$ & \ldots  & \dots & \ldots & \phn267$^{b}$ \\
01045-SMM2 &  \ldots & 32$^{a}$ & \ldots  & \dots & \ldots & \phn432$^{b}$ \\
&&&&&&\\
\hline
\hline
\end{tabular}
\flushleft
$^a$ Assumed \tdust.\\
$^b$ Mass calculated from S$_{850}$ assuming the \tdust\ given 
in the table and \\ dust emissivity index $\beta=2$.  
\end{table*}

Except for the sub-mm cores in IRAS 05345+3556, the greybody models
produce reasonable fits to the SEDs of the sub-mm cores in IRAS
05271+3059, 20222+3541 and 20406+4555. The SED models for the cores in
IRAS 05345+3556 do not reproduce the 450~\micron\ fluxes. A somewhat
less constrained two dust temperature greybody model fit to the SED of
the sub-mm cores in IRAS 05345+3556 yields better fits. This indicates
that additional observations between 100 and 450~\micron\ are
essential to fully constrain the fits to the SEDs.

\section{Dust Properties from the SED fits}
\label{sec_dustprop}

The temperatures of the cold components for all the sub-mm cores are
$\sim 32\pm5$~K and the dust emissivity indices vary between
0.9 and 2.5. While the sub-mm cores within IRAS 05271+3059 show very
different dust emissivity indices ($\beta$; Table~\ref{greytab}), the
three sub-mm cores in IRAS 20406+4555 show almost identical values of
$\beta$.

The dust grain emissivity index is crucial in determining many source
parameters, including the dust temperature, as well as providing
information about the grain structure and the dielectric
characteristics of the grains \citep{schwartz1982,hildebrand1983}.
Knowledge of $\beta$ improves estimates of the dust opacity and
subsequently of the dust mass of the star forming clouds. Various
estimates of $\beta$ from both laboratory and theoretical models have
been proposed and so far there is little consensus between these
studies.  \citet{draine1984} derived $\beta \sim 2$ for 40$\leq
\lambda \leq 1000$~\micron, using the available laboratory data to
measure optical constants of a mixture of naked graphite and silicate
grains. A dust grain emissivity index of $\sim 2$ has also been
supported by the calculations of
\citet{knacke1973,mathis1989,kruegel1994}.  \citet{miyake1993} however
derived values of $\beta$ in excess of 2 for large grain sizes, while
\citet{wright1987} calculated 0.6$\leq \beta\leq1.4$ for fractal
grains. \citet{aannestad1975} predicted $\beta$ up to 3.5 for olivine,
fused quartz, and lunar rock grains covered with ice mantles.
\citet{mennella1995} concluded that more evolved cores have a higher
emissivity index ($\beta$).

Observationally there is an equally large scatter in the derived
values of dust emissivity index, $\beta$, as in theoretical studies.
There are several instances of observations of circumstellar disks
around young PMS stars \citep{beckwith1991,sandell2001b,natta2004},
low mass protostars \citep{hogerheijde2000,stark2004} as well as high
mass protostars \citep{sandell2000,sandell2004} for which $\beta<2$
were derived.  Larger values of $\beta$, i.e., in excess of 2 have
also been reported \citep{kuan1996,goldsmith1997,lis1998}.  We note
here that the derived values of $\beta$ is always to some extent
affected by the temperature and density variations as long as there is
some optically thick portion of dust and detailed radiation transfer
models can achieve a good fit to observations using $\beta=2$
\citep{mueller2002,shirley2002}.  In addition,
uncertainties and artifacts due to ignored error-lobe pickup,
calibration uncertainties, optical thickness of \HII\ regions, 
unequal beamsizes at the two wavelengths under consideration etc. have
often contributed to observationally determined higher values of
$\beta$.

The values of $\beta$ derived from the SCUBA observations of the outer
Galaxy sources lie within the limits of the different values seen in
the inner Galaxy sources.  Based on the present estimates we conclude
that the overall dust properties shown by the outer Galaxy sources, do
not differ significantly from the properties of the inner Galaxy star
forming regions, and show no indication of the substantially higher
values of $\beta$ as found in a few extreme regions in the Galactic
Center environment. The protostellar sources observed here are rather
distant, hence the dust emission is dominated by the extended dense
cloud envelopes surrounding the young high mass protostellar objects,
rather than being dominated by the emission from the accretion disk,
which are always found to show a lower apparent $\beta$.

\section{Radial Profiles }
\label{sec_radprof}

\subsection{Fits to the radial intensity profiles}

\begin{figure}
\includegraphics[width=9.0cm,angle=0]{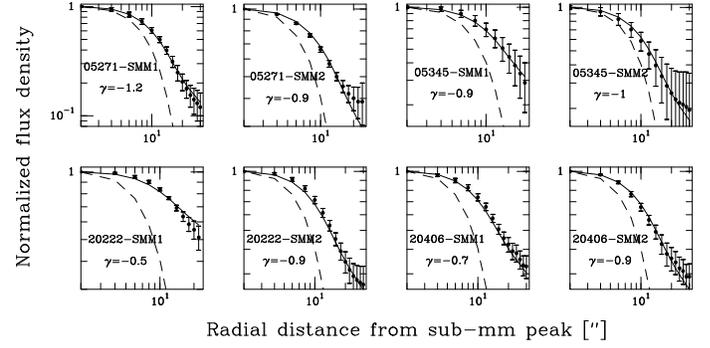}
\caption{Normalized, azimuthally averaged radial profiles for the
intensities of the sub-mm cores at 850~\micron\ ({\em points}). {\em
Solid line} shows the best fit power-law model corresponding to the
index $\gamma$.  The dashed line shows the model beam profile.  The
errorbars indicate the rms values within the annuli.
\label{fig_radprof}}
\end{figure}

Figure~\ref{fig_radprof} shows the radial intensity profiles at
850~\micron\ of all the sub-mm cores with the exception of 05345-SMM3
and 20406-SMM3.  These radial profiles have been derived by averaging
all datapoints within a 2\arcsec\ wide circular annulus at a radius
$r$ from the center of the core, with $r$ ranging from 0 to 30\arcsec.
The observations were done with a chop throw of 2\arcmin\ in the
azimuth.  Based on the derived angular sizes of the sub-mm cores we
conclude that the effect of chopping on the derived radial intensity
profiles is negligible.

We have created model intensity distribution by assuming an
analytical two-dimensional power-law intensity distribution of the form
$I(r)\propto \theta^\gamma$ and convolving it with the synthetic
beam profile at 850~\micron\ which we used to deconvolve the intensity
maps. We derived the model radial intensity profile by azimuthally
averaging exactly in the same manner in which we have created the
observed intensity profile. We have performed a $\chi^2$ fitting of a
family of such model profiles corresponding to different power law
indices to the observed model profile.  Keeping in mind the angular
resolution of the 850~\micron\ images we have restricted the fitting
procedure to radii $8$\arcsec$<r<30$\arcsec, so as to have a reliable
estimate of the nature of the intensity distribution at spatial scales
unaffected by the resolution of the observations.
Figure~\ref{fig_radprof} shows the resultant fits and
Table~\ref{tab_index} presents the fitted values of  $\gamma$.

For most of the cores it is possible to derive acceptable fits to the
radial intensity profiles of the outer envelopes ($r>8$\arcsec) using
a single power law. The fits, though not constrained by the inner
regions, are reasonably good at radii $<$8\arcsec\ as well and we
derive  $\gamma=-0.9\pm0.3$ for the 8 sub-mm cores.

\subsection{Density Profiles of the sub-mm cores}

Here we consider a model with a spherically symmetric envelope and use
the intensity distribution of the dust emission in the SCUBA images at
850~\micron\ to derive a first estimate of the radial density profile
in the envelope. The density profile is a power law, $n \propto r^p$,
the value of $p$ derived using this method depending upon the assumed
temperature distribution.

For optically thin dust emission, assuming the Rayleigh-Jeans
approximation to be valid, spherical cores and temperature and density
distribution following a power law, the intensity index $\gamma$
depends to first order on the density index $p$ and the temperature
index $q$ ($T \propto r^{q}$) through \citep{motte2001,adams1991}:

\begin{equation}
\gamma = 1 - p - q 
\end{equation}

\begin{table}
\caption{Power law indices for radial intensity \& density profile
\label{tab_index}}
\begin{center}
\begin{tabular}{ccc}
\hline
\hline
Source & $\gamma$ & $p$\\
\hline 
05271-SMM1 &  -1.2 & -1.8 \\
05271-SMM2 &  -0.9 & -1.5 \\
05345-SMM1 &  -0.9 & -1.5 \\
05345-SMM2 &  -1.0 & -1.6 \\
20222-SMM1 &  -0.5 & -1.1 \\
20222-SMM2 &  -0.9 & -1.5 \\
20406-SMM1 &  -0.7 & -1.3 \\
20406-SMM2 &  -0.9 & -1.5 \\
\hline
\end{tabular}
\end{center}
\end{table}

Thus to derive the density index $p$ it is necessary to have some
knowledge about the temperature distribution. Based on the grey-body
fitting of the SEDs  it is apparent that the sub-mm cores are
centrally heated (accretion and/or stellar burning) so that the
temperature decreases with increasing distance from the center.
Theoretical calculations of massive star forming regions  suggest 
that steeper indices for the temperature distribution are expected in
the inner regions ($<$2000 AU).  However beyond, the inner 2000 AU,
the temperature distribution flattens again and runs asymptotically
into a $r^{-0.4}$ distribution \citep{vandertak2000}.  For the large
distances to our sources, the typical resolution of the 850~\micron\
maps are well beyond 20,000 AU. This implies that we can safely assume
the asymptotic solution of $q=0.4$ for our sources. Thus, for parts of
the sub-mm cores at radii larger than 20,000 AU we derive density
indices varying between -1.1 and -1.8, with most cores showing indices
of -1.5 (Table~\ref{tab_index}).

Analytic and numerical studies show that the mass infall rate during
the protostellar phase is dependent on the radial density profile at
the onset of collapse, and on the equation of state of the material.
\citet{motte2001} studied isolated and clustered protostellar
envelopes, and found that isolated protostellar envelopes exhibit
density profiles like $n \propto r^{-2}$, as predicted by the standard
model of star formation by \citet{shu1987}. In contrast, protostellar
envelopes in clusters are found to be induced in compact condensations
resembling more finite-sized Bonnor-Ebert spheres than singular
isothermal spheres, which suggests that dynamical protostellar models
are more appropriate \citep{motte2001,whitworth1985}. ISO based
mid-infrared absorption studies of low-mass prestellar cores revealed
a flattening of the inner parts of the radial profiles as well as a
steepening of the profiles farther outside \citep{bacmann2000}. In the
high mass regime studies of ultracompact \HII\ regions and/or hot
cores indicate that the overall density profiles are more like $n
\propto r^p$ with $-1.0<p<-1.5$ as predicted by logatropic equations
of state \citep{hatchell2000,vandertak2000}.  Based on the study of a
large sample of high mass protostellar objects (HMPOs)
\citet{beuther2002} derived a much wider range of density indices $n
\sim r^{-1.6\pm0.5}$. A possible physical interpretation of the
flattened density profile for massive objects  could be that in
low-mass objects the support against gravitational collapse is
thermal, while for massive objects nonthermal support plays a major
role \citep{myers1992,mclaughlin1996}.

Based on the luminosity estimates (Table~\ref{greytab}) the sources we
consider here are all intermediate to massive star forming regions,
some of them having \HII\ regions associated with them
(Fig.~\ref{fig_intmap}). The derived value of density indices for the
8 sub-mm cores is $-1.5\pm0.3$ and is thus consistent with
observations of massive star forming regions and agree with
theoretical models with significant non-thermal support against
gravitational collapse.

\section{Summary}

The high angular resolution (15\arcsec) sub-mm dust continuum maps of
the outer Galaxy IRAS point sources show that all of them contain
multiple emission peaks. Only four of the fourteen detected sub-mm
sources have associated mid-infrared emission as seen from the MSX
point source catalog and/or Spitzer IRAC and MIPS observations. SEDs
for individual sub-mm cores between 60 and 850~\micron\ have been
generated by apportioning the unresolved IRAS fluxes at 60 and
100~\micron\ in the same ratio as the observed 450~\micron\ fluxes.
Single temperature greybody models were fitted to the observed SEDs of
the sub-mm cores to derive the dust temperature, emissivity index
($\beta$) and optical depth at $\tau_{250}$. The derived dust
temperature of the sub-mm cores is 32$\pm$5~K and the emissivity index
ranges between 0.9 and 2.5. The two sub-mm cores in IRAS 05271+3059
show largely different emissivity indices.  In spite of the known
radial gradients in metallicity within the Milky Way, the dust
properties derived here do not show any significant difference from
the inner Galaxy star forming regions.

An analysis of the radial intensity profiles (for $r>$ 8\arcsec) at
850~\micron\ for the sub-mm cores shows that it is possible to fit the
profiles reasonably well using a single power law with indices
-0.9$\pm0.3$. Deriving density distributions from the radial indices
we get density power laws -1.5$\pm$0.3, with most cores showing an
index of -1.5.  This density profile is similar to the profile
observed in many massive star forming regions in the inner Galaxy and
generally agree with models of star formation which invoke non-thermal
support against gravitational collapse
\citep{myers1992,mclaughlin1996}.

Thus, except for their isolated occurence in the outer Galaxy, in most
other respects the sources studied here look very simiilar to star
formation cores in the inner parts of the Galaxy which have similar
range of masses.

\section*{Acknowledgments}

The James Clerk Maxwell Telescope is operated by The Joint Astronomy
Centre on behalf of the Science and Technology Facilities Council of
the United Kingdom, the Netherlands Organisation for Scientific
Research, and the National Research Council of Canada.  Data in this
paper were obtained under the program ID M04BN04. This material is
based upon work supported by the Deutsche Forschungs Gemeinschaft
(DFG) via grant SFB494 and the National Science Foundation under Grant
No.  AST-0228974.  This research has made use of NASA's Astrophysics
Data System.


\end{document}